\documentclass[a4paper,11pt]{article}
\pdfoutput=1 

\usepackage{jheppub} 
\usepackage{float}
\usepackage[OT1]{fontenc} 
\usepackage{subcaption}

\title{\boldmath Cluster Scanning: a novel approach to resonance searches}

\author[a,b]{I. Oleksiyuk,}
\author[a]{J. A. Raine,}
\author[c]{M. Kr\"amer,}
\author[b]{S. Voloshynovskiy,}
\author[a]{T. Golling}

\affiliation[a]{Département de Physique Nucléaire et Corpusculaire, University of Geneva, 1211 Geneva, Switzerland}
\affiliation[b]{Centre Universitaire d'Informatique, University of Geneva, 1211 Geneva, Switzerland}
\affiliation[c]{Institute for Theoretical Particle Physics and Cosmology, RWTH Aachen University, 52074 Aachen, Germany}

\emailAdd{ivan.oleksiyuk@unige.ch}
\emailAdd{john.raine@unige.ch}
\emailAdd{mkraemer@physik.rwth-aachen.de}
\emailAdd{svyatoslav.voloshynovskyy@unige.ch}
\emailAdd{tobias.golling@unige.ch}

\abstract{We propose a new model-independent method for new physics searches called Cluster Scanning. It uses the k-means algorithm to perform clustering in the space of low-level event or jet observables, and separates potentially anomalous clusters to construct a signal-enriched region. The spectra of a selected observable (e.g. invariant mass) in these two regions are then used to determine whether a resonant signal is present. A pseudo-analysis on the LHC Olympics dataset with a $Z'$ resonance shows that Cluster Scanning outperforms the widely used 4-parameter functional background fitting procedures, reducing the number of signal events needed to reach a $3\sigma$ significant excess by a factor of 0.61. Emphasis is placed on the speed of the method, which allows the test statistic to be calibrated on synthetic data.}

\begin{document} 
\maketitle
\flushbottom

\section{Introduction}
The Standard Model (SM) is the current apex of theoretical physics, describing the electromagnetic, weak and strong interactions with unparalleled precision. Unfortunately, it is still far from complete, as several phenomena remain unexplained. In order to create a ``theory of everything", one would not only need to combine the SM with general relativity, but also provide an explanation for many other issues, including the existence of neutrino masses, the origin of the matter-antimatter asymmetry, and most importantly, the origin of dark matter. To solve these problems, researchers are collaborating to formalise new theories, design, build and carry out new experiments, as well as simulate and analyze research data.

One of the most renowned experimental facilities, the Large Hadron Collider (LHC) was constructed with the purpose of testing the SM in the high energy regime. The last elementary particle predicted by the SM, the Higgs boson, was discovered in 2012 \cite{Aad_2012, Chatrchyan_2012}. Since then, LHC research has shifted towards precision measurements and searches for beyond the Standard Model~(BSM) effects. 

Many extensions of the SM imply the existence of as yet undiscovered massive particles, often associated with proposed new symmetry groups. 
If a new particle has a narrow decay width, the straightforward method is to search for a resonant peak in the spectrum of a mass-like observable, such as the invariant mass of a dijet event.
However, such a bump hunt is not completely free of assumptions.
Often complex analytical functions need to be chosen to model the background distribution, with the possibility to introduce spurious signals and varying sensitivity under the assumption of different functional forms.
Furthermore, additional observables or fiducial cuts need to be chosen and optimised to enhance sensitivity in the case where potential signal yields are low, causing searches to become more model-specific. To broaden the range of signal models covered by searches one may employ the model-unspecific search steategies \cite{Knuteson:2001dq, CDF:2007ykt, CMS:2020zjg}.

Over the past decade, machine learning-based algorithms have become increasingly popular for solving a multitude of problems. Deep learning, in particular, has gained popularity for various tasks, with large neural networks being utilised. For example, many methods were implemented to perform anomaly detection~(AD) tasks in various industries.
Some of these AD methods have been repurposed and extended to support BSM searches~\cite{Collins:2019jip, Collins:2018epr,DAgnolo:2018cun,DAgnolo:2019vbw,Farina:2018fyg,Heimel:2018mkt,Roy:2019jae,Cerri:2018anq,Blance:2019ibf,Hajer:2018kqm,DeSimone:2018efk,Mullin:2019mmh,Dillon:2019cqt,Aguilar-Saavedra:2017rzt,Romao:2019dvs,Romao:2020ojy,Amram:2020ykb,Cheng:2020dal,Khosa:2020qrz,Thaprasop:2020mzp,Alexander:2020mbx,Mikuni:2020qds,vanBeekveld:2020txa,Park:2020pak,Faroughy:2020gas,massiveissue,Chakravarti:2021svb,Batson:2021agz,Blance:2021gcs,Bortolato:2021zic,Collins:2021nxn,Dillon:2021nxw,Finke:2021sdf,Shih:2021kbt,Atkinson:2021nlt,Kahn:2021drv,Dorigo:2021iyy,Caron:2021wmq,Govorkova:2021hqu,Kasieczka:2021tew,Volkovich:2021txe,Govorkova:2021utb,Ostdiek:2021bem,Fraser:2021lxm,Jawahar:2021vyu,Herrero-Garcia:2021goa,Aguilar-Saavedra:2021utu,Tombs:2021wae,Lester:2021aks,Mikuni:2021nwn,Chekanov:2021pus,dAgnolo:2021aun,Canelli:2021aps,Ngairangbam:2021yma,Bradshaw:2022qev,Aguilar-Saavedra:2022ejy,Aguilar-Saavedra:2023pde,Buss:2022lxw,Alvi:2022fkk,Dillon:2022tmm,Birman:2022xzu,Letizia:2022xbe,Fanelli:2022xwl,Finke:2022lsu,Verheyen:2022tov,Dillon:2022mkq,Caron:2022wrw,Park:2022zov,Kamenik:2022qxs,Kasieczka:2022naq,Araz:2022zxk,Schuhmacher:2023pro,Roche:2023int,Vaslin:2023lig,ATLAS:2023azi,Chekanov:2023uot,CMSECAL:2023fvz,Bickendorf:2023nej,Freytsis:2023cjr,Metodiev:2023izu}
(see Refs.~\cite{karagiorgi2022machine,LHCO,aarrestad2022darkMachines,Belis:2023mqs} for a comparison of various ML assisted BSM methods and Refs.~\cite{collins2021compare,Golling:2023yjq} for a comparison of weakly supervised and unsupervised approaches).
The ATLAS collaboration produced the first experimental results for such searches applied to experimental data using weakly supervised methods~\cite{atlas2020cwolaExp} and unsupervised ML anomaly detection methods~\cite{ATLAS:2023azi,atlas2023twoBody}. 
However, these efforts have not observed any significant deviations from the SM expectation.

Many AD approaches rely on the assumption that any new signal would form a set of outliers.
However, in a bump hunt the assumption is instead that any new signal would be localised in some feature space, in particular in an invariant mass spectrum.
Weakly supervised approaches, on the other hand, aim to enhance the sensitivity by applying a cut on a classifier trained directly on the data.
However, in both instances the same bump hunt restrictions apply with either functional forms or input observables impacting the sensitivity to a model.

In this work we introduce a new data-driven method, Cluster Scanning~(CS), which builds on the foundations of the bump hunt but addresses several limitations.
By leveraging more information from the event CS is able to enhance sensitivity to potential signals without enforcing any model specific assumptions, and can also provide a direct estimate of the background distribution.
The proposed approach complements existing techniques and is designed to be computationally efficient.

The paper is structured as follows.
In Section~\ref{sec:data}, we briefly describe the LHCO R\&D dataset~\cite{LHCOlympics}, commonly used to benchmark the performance of anomaly detection techniques, and introduce our data preprocessing steps. Section~\ref{sec:method} touches on the general topic of bump-hunting strategies in the literature, introduces the novel CS method, and discusses similarities and differences between them. In Section~\ref{sec:results} we provide the results of applying CS in an anomaly search. Finally, we draw conclusions in Section~\ref{sec:conclusion}.

\section{Dataset}
\label{sec:data}
The LHCO R\&D dataset consists of one million background Standard Model dijet events (also subsequently referred to as QCD) and 100\,000 signal BSM $Z'\rightarrow XY$ events, where massive particles with $m_X=500$\,GeV and $m_Y=100$\,GeV decay into quark-antiquark pairs. The resonance itself has a mass of $m_{Z'}=3.5$\,GeV. This anomaly model is discussed in detail in Ref.~\cite{Kim_2020}. 

All the events were produced using \textsc{Pythia}~8.219~\cite{Sjostrand:2014zea} and \textsc{Delphes}~3.4.1~\cite{de_Favereau_2014, Mertens_2015, Selvaggi_2014} using default settings. The jets were clustered using an anti-$k_T$ algorithm \cite{Cacciari_2008} with $R=1$ using \textsc{FastJet} \cite{Cacciari_2012} with a \texttt{python} interface provided through the \texttt{pyjet} library in \textsc{Scikit-HEP} \cite{Rodrigues:2020syo}.
Jets are required to have $p_T>1.2$\,TeV and fall within $|\eta|<2.5$.

\subsection{Jet images}

In addition to the di-jet invariant mass ($m_{jj}$) of the event, used in a bump hunt, we extract additional information from the image representations of the two jets.
This allows for a more model agnostic approach than selecting specific jet substructure observables.
The jet images are processed following a prescription similar to that used in Ref.~\cite{Jet_images,Kasieczka_2019, Macaluso_2018, Heimel:2018mkt} from the $\eta$, $\phi$ and $p_\mathrm{T}$ of the jet constituents.
Individual jet images are centred, rotated, and flipped in order to provide a consistent input to a convolutional neural network, reducing the number of symmetries the ML method would need to learn.

\begin{figure}[h]
    \centering
    \includegraphics[width=1\linewidth]{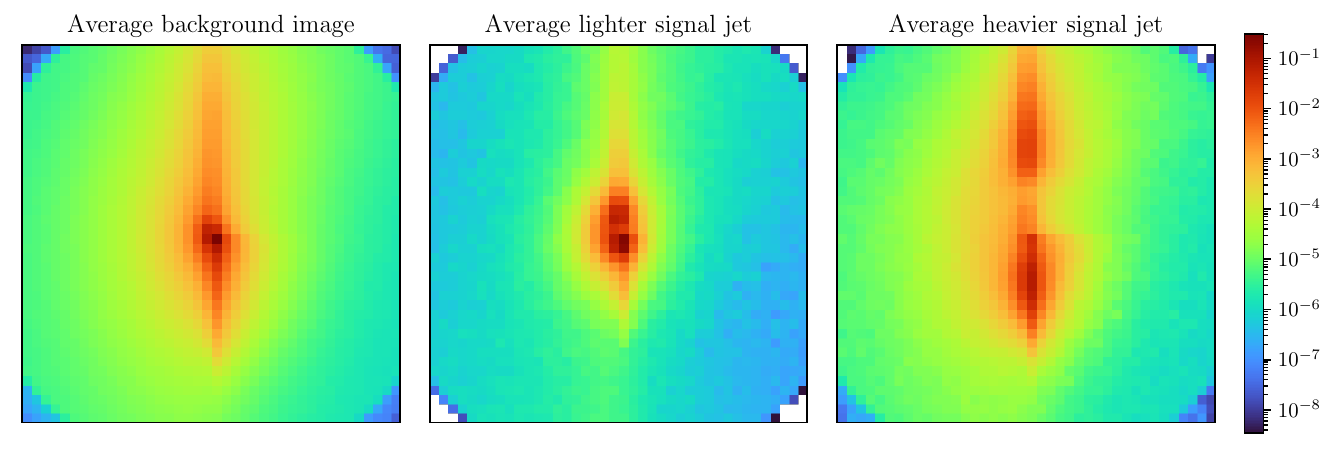}
    \caption{From left to right: average of all 2M available QCD jet images, average image of all 100K lighter jets in a $Z'$ event and average image of all 100K heavier jets in a $Z'$ event before smearing and pixel scaling.}
    \label{fig:mean_QCD}
\end{figure}

The jet images are cropped to $[-0.8, 0.8]\times[-0.8, 0.8]$ in $\eta-\phi$ space relative to the jet centre, binned with a $40\times40$ pixel grid, and normalised such that the sum of all pixels is equal to one.
Fig.~\ref{fig:mean_QCD} shows the average jet images for QCD background, and the separate averages of all lighter (mostly $Y$) and heavier (mostly $X$) jets in each $Z'$ event.

Despite being used in many applications, the jet image representation has two main drawbacks, namely the sparsity of non-zero pixels (see app. \ref{app:sparse}) and the imbalance in the magnitudes of their intensities.
This is particularly problematic for approaches that depend on the $L_2$ (Euclidean) distance. We address both of these problems with the solutions introduced in Refs.~\cite{2021AEanomalies, Buss:2022lxw}. 

To take the soft constituents into account, which have intensities orders of magnitudes lower than hard constituents, we apply a non-linear scaling to all pixels of $I_{ij} \rightarrow I_{ij}^\gamma$. To address sparsity we convolve (smear) the whole image with a two-dimensional Gaussian kernel with an isotropic standard deviation $\sigma_k$. 
We find that using a value of $\gamma=0.5$ for the pixel scaling alongside $\sigma_k=1$ for the Gaussian kernel provides a adequate solution to both issues without excessive impact on the structure of the jets.

\section{Method}
\label{sec:method}
\subsection{Bump hunt}
\label{sub:bump_hunt}

The bump hunt approach is a standard method used to search for excesses over a non-resonant background in HEP (high-energy physics) data. This method usually follows four main steps that we briefly discuss below. Each of these steps is a complex topic in  itself with several different approaches in the literature, thus for our study, we choose only simplified and basic approaches.

\subsubsection{Signal enrichment}
\label{sub:sig_en}
Signal enrichment, in general, refers to the selection of a subset of experimental data in such a manner that the fraction of signal events in it is increased compared to the initial sample. Most often, this is done by cutting out a region of the observable space where the signal is expected to be abundant compared to the rest of the space, typically using a theoretical model of the signal of interest.

These approaches, despite being sensitive to specific signal processes, make the search less model-agnostic and are ill-suited for general anomaly detection searches. Alternatively, one can hope to define a signal-rich region of the experimental data using a plethora of unsupervised ML (machine learning) techniques, which are expected to provide enhanced sensitivity over a wider range of potential signal processes.

In our particular example of LHCO data, we choose to explore a wide, smoothly falling region of the spectrum of dijet events with invariant dijet mass $m_{jj}$ from 3000\,GeV to 4600\,GeV. We choose this lower bound to avoid the turn on curve of the mass distribution, resulting from the jet trigger, and the upper bound is selected to remain in a region with relatively high statistics, so that we work in the region where the fit functions from Subsection~\ref{subsec:bkg} are applicable. This interval contains, in total, around 380,000 QCD events and nearly all $Z'$ events. We divide this region into 16 non-intersecting bins with 100\,GeV width each, as in Ref. \cite{cathode, curtains}.

\subsubsection{Background estimation}
\label{subsec:bkg}
To perform a hypothesis test, one must first postulate a null hypothesis, which in counting experiments takes form of the expected background coming from the Standard Model processes. Often the background prediction relies on a theoretical basis to calculate the cross sections of the hard process and a simulation to account for detector response and measurement uncertainties. 
Still there are a number of searches where theory and simulation cannot provide a reliable background estimate. 
In these cases the background has to be estimated from the data itself in an empiric manner, using some general assumptions. 

In dijet-like searches a background is often estimated by fitting a function of the form
\begin{equation}
\label{eq:function}
f(x)=p_1(1-x)^{p_2}x^{p_3+p_4\ln(x)+p_5\ln(x^2)}
\end{equation}
to a smoothly falling part of the dijet mass distribution~\cite{CMS:2022eud, 4par_ATLAS_09_04_2015,3par_ATLAS_03_03_2016,4par_CMS_2017, Khachatryan_2016, Khachatryan_2016_2, 4par_ATLAS_20_04_2011, 4par_CMS_2015, PhysRevD.87.114015, 4par_CMS_2013, 4par_ATLAS_2012, 4par_ATLAS_2013, 4par_CMS_2011, 4par_CDF_2009, Aad_2011, PhysRevLett.105.211801,5-parameter}, where $x=m_{jj}/\sqrt{s}$.
This function is referred to as the ``$n$-parameter dijet fit function'', where $n$ is the number of nonzero free parameters $p_i$ used in the function. 
Despite being a good fit to the simulated data, this functional form is still an empirical assumption and thus is subjects to a systematic error. 
Furthermore, after applying some selection criteria on the events which could be correlated with $m_{jj}$, this function may no longer well describe the resulting distribution.

More advanced methods of fitting, such as the Sliding Window Fit (SWIFT) \cite{SWIFT} and the ABCD method used in \cite{Alison:2229034, PhysRevD.103.035021} are other methods that reduce the assumption of a functional form but introduce their own assumptions instead. 
However, due to the simplicity and wide use of the $n$-parameter fit function, we choose to use global 3-parameter and 4-parameter function fits as the benchmark analysis strategy.
Further details of the (pseudo-)analysis on the LHCO R\&D data performed using these background estimates are given in Appendix~\ref{app:global_fits}. 
To access the upper bound on the performance of all background estimation methods, we use the underlying background distribution as an idealised fit, i.e.\ a fit with no systematic error. The (pseudo-)analysis using this is also described in Appendix~\ref{app:global_fits}.

\subsubsection{Test Statistic definition and calibration}
\label{sub:test_stat}
There are several ways to calculate a global test statistic for two spectra.
In HEP one of the more popular tests in model agnostic searches, called BumpHunter~\cite{vaslin2023pybumphunter}, relies on the maximal local significance~(MLS) as the test statistic, where it is computed using a range of different windows over the spectrum. 
One of the benefits of the MLS test statistic is its simplicity and that it is well suited for signals that give rise to narrow, localised resonances.
Here the MLS is applied to the binned $m_{jj}$ distributions of the data. Given a set $\mathcal{B}=\{b_1, ..., b_{n_{\rm bins}}\}$ of non-intersecting bins with $N_{\text{sig+bkg},b}$ events or jets from the signal-rich (experimental) distribution and $N_{\text{bkg},b}$ events or jets from the background estimation, the MLS can be written as
\begin{equation}
\label{eq:poisson}
\begin{split}
\text{MLS} = \underset{b\in \mathcal{B}}{\text{max}}Z_b = \underset{b\in \mathcal{B}}{\text{max}}(\text{CDF}^{-1}_{\mathcal{N}(0, 1)}(\text{CDF}_{Poisson(N_{\text{bkg},b})}(N_{\text{sig+bkg},b})))\,,
\end{split}
\end{equation}
where CDF is the cumulative density function of the respective distribution. In equation \ref{eq:poisson} only overdensities are taken into account, i.e. $Z_b>0$ only for $N_{\text{sig+bkg},b}>N_{\text{bkg},b}$ as we are searching for a resonance. 

For bins with $N_{\text{sig+bkg}/\text{bkg},b}\gg 1$ one can approximate the Poisson distribution with a normal distribution $\mathcal{N}(N_{\text{sig+bkg}/\text{bkg},b}, \sqrt{N_{\text{sig+bkg}/\text{bkg},b}})$. 
Equation \ref{eq:poisson} then reduces to a much simpler form 

\begin{equation}
\label{eq:poisson2}
\begin{split}
\text{MLS} = \underset{b\in \mathcal{B}}{\text{max}}Z_b = \underset{b\in \mathcal{B}}{\text{max}}\frac{N_{\text{sig+bkg},b}-N_{\text{bkg},b}}{\sqrt{N_{\text{bkg},b}}}\,.
\end{split}
\end{equation}

Although some test statistics, like $\chi^2$, have well-known distributions, other more unusual test statistics, like the BumpHunter test statistic, require calibration. This is commonly done by modelling its distribution using Monte Carlo simulation.

Moreover, as systematic uncertainties arise from the definition of a signal region selection and the background estimate, this calibration should be performed even in the case where the distribution is known a priori.
The calibration for the BumpHunter test statistic is performed in Ref. \cite{vaslin2023pybumphunter} by running pseudo-experiments in which the counts in each bin are varied according to Poisson's law. This can be extended to higher dimensions by resampling the background events with bootstrapping.
By calculating the test statistic for each of our bootstrapped background-only pseudo-experiments, we obtain the distribution of the test statistic in the background-only hypothesis.
To ensure good modelling of the tail of the test statistic distribution, which corresponds to large significance values in the presence of signal, a large number of pseudo-experiments is required.

\subsubsection{Significance evaluation}
\label{sub:significance_ev}
To obtain a calibrated $p$-value for a given value of the test statistic $t$, one counts the number of background only pseudo-experiments exceeding this value $N_{>t}$ and divides it by the total number of pseudo-experiments done, $N_{tot}$.

The (one-sided) significance is computed using the inverse cumulative density function of the normal distribution $Z=\mathrm{CDF}^{-1}_{\mathcal{N}(0, 1)}(1-p\text{-value})$. 

In the case of $N_{>t}=0$ arising from the limited number of pseudo-experiments, we instead set a lower bound:
\begin{equation}
    p<\frac{1}{N_{tot}},\; Z>\mathrm{CDF}^{-1}_{\mathcal{N}(0, 1)}(1-\frac{1}{N_{tot}})\,.
\end{equation}

For every experiment with added signal events, we still bootstrap the background (for consistency) and combine it with a given number of signal events chosen at random from 100,000 signal events (around 5\% of events fall outside of the evaluation region). 
Due to statistical fluctuations we also perform several pseudo-experiments in the signal enriched case in order to obtain a robust estimate of the significance for each level of signal doping.

\subsection{Cluster scanning}

\label{sub:algo}
In this section we present a novel approach called Cluster Scanning, which follows the same bump hunting scheme, but relies on a distinct set of assumptions than the commonly employed methods and thus has several favorable characteristics. Our approach can be divided into several key steps given below, with the hyperparameters chosen in order to search for narrow resonances in the $m_{jj}$ spectrum of the LHCO R\&D data. The motivation for these hyperparameters in each step and the argumentation on how to choose them for a different application case is given in App.~\ref{app:hyper}.

\paragraph{Training region selection:} We select a narrow $m_{jj}$ window [3000, 3100]\,GeV for training of the k-means algorithm. This window contains 56,486 original background events. In this publication, we focus on relatively small signal injections that include only 5\% or less of the total number of Z' signals available. Therefore the training region is expected to contain 89 signal events or less, which can be regarded as negligible (proof given in App.~\ref{app:signal_in_training}). In App.~\ref{app:signal_in_training} we show an improvement in performance in case the training region matches the resonant peak and thus has a larger portion of signals events involved in clustering. However, in an actual analysis the position of the peak will be unknown, thus we choose to discuss a more representative case given here, when the training region happens to be in the tail of the signal peak and thus has a negligible number of signal events.

\paragraph{K-means Clustering:}
We apply a mini-batch $k$-means clustering algorithm with \mbox{$k=50$} implemented in the \textsc{scikit-learn} \cite{scikit-learn} library, with a batch size of 2048 on the set containing jet images of the leading two jets from each event in this $m_{jj}$ window. The mini-batch implementation is chosen due to its computational speed. The seeding of the cluster centroids is performed using the \textsc{k-means++} prescription described and motivated in Ref.~\cite{kmeans}.

\begin{figure}[h]
  \centering
  \includegraphics[width=0.789\textwidth]{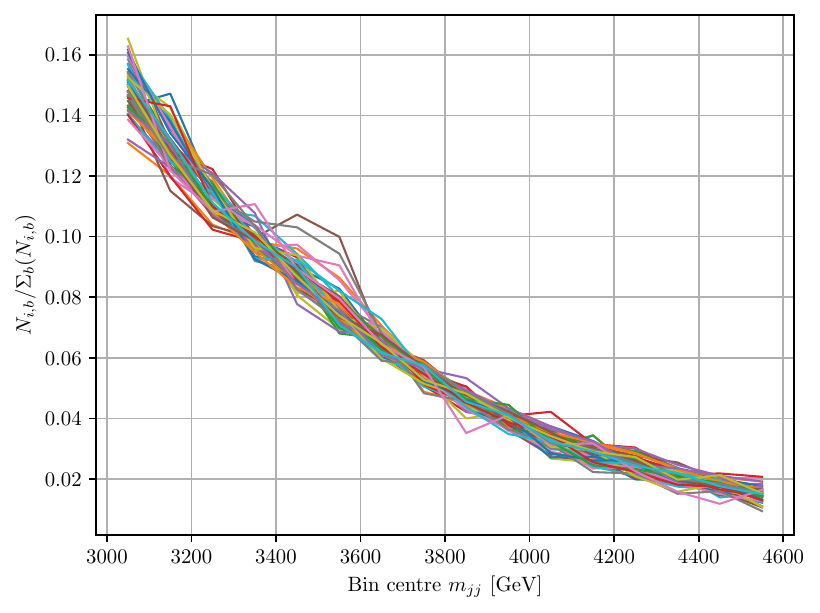}
  \caption{The $m_{jj}$ distributions for the jets in each of the 50 clusters, each normalised to unity. Here, 5,000 signal events have been injected into the evaluation dataset, which corresponds to 5\% of the total available signal events.}
  \label{fig:50_clusters_nrm}
\end{figure}

\paragraph{Cluster Spectra:}
After performing the fit of $k$-centroids to the data in the training region, we fix the centroid positions and evaluate how many jet images from each of the 16 $m_{jj}$ bins of the evaluation region, defined in Subsection~\ref{sub:sig_en} (it includes training region as one of the bins), fall into each of the $k$ clusters $N_{i, b}$ where $i\in\{1 ... k\},\,b\in\{1 ... n_{bins}\}$.
Fig.~\ref{fig:50_clusters_nrm} shows the resulting 50 normalised cluster spectra $N_{i, b}/\sum_b(N_{i, b})$ for one pseudo-experiment with signal injection. 

\begin{figure}[h]
  \centering
  \includegraphics[width=0.789\textwidth]{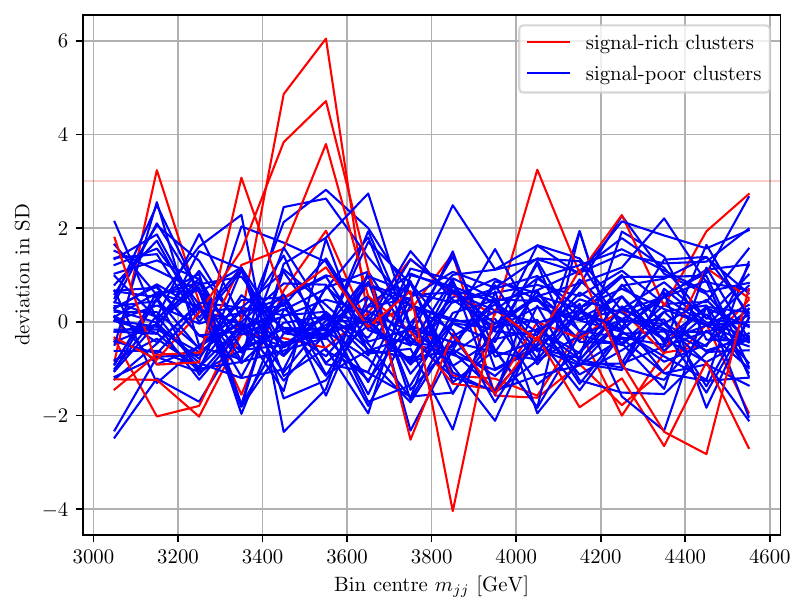}
  \caption{Spectra in Fig.~\ref{fig:50_clusters_nrm} standardised over clusters in each bin. Potentially signal-rich cluster spectra are shown in red.}
  \label{fig:norm_subtr_std}
\end{figure}

\paragraph{Per bin standardisation:}
We note that in each bin the normalised cluster spectra follow an approximately normal distribution with several outliers from the anomalous clusters (see discussion in App.~\ref{app:gaussian_dist}). Therefore we standardise the normalised cluster spectra in each bin using outlier robust estimators (described in App.~\ref{app:robust_estimators}) for mean and standard deviation with an outlier factor of 0.2. Here we make the assumption that the majority of the signal is located in a small number of clusters, and the rest of the clusters are signal depleted.  
Figure~\ref{fig:norm_subtr_std} shows the cluster spectra from Fig.~\ref{fig:50_clusters_nrm} after normalising with the outlier robust estimator.

\paragraph{Selecting anomalous clusters:}
Utilising the assumption that the signal is localised in $m_{jj}$, we select potentially signal-rich cluster spectra as those with a deviation of more than a threshold value of $\theta=3$ standard deviations from the robust mean in the positive direction as we are only interested in a resonance leading to excess of events. The rest of the clusters are labelled as signal-depleted. The threshold and the selected signal-rich clusters are shown in Fig.~\ref{fig:norm_subtr_std} in red.

\begin{figure}[h]
  \centering
  \includegraphics[width=0.79\textwidth]{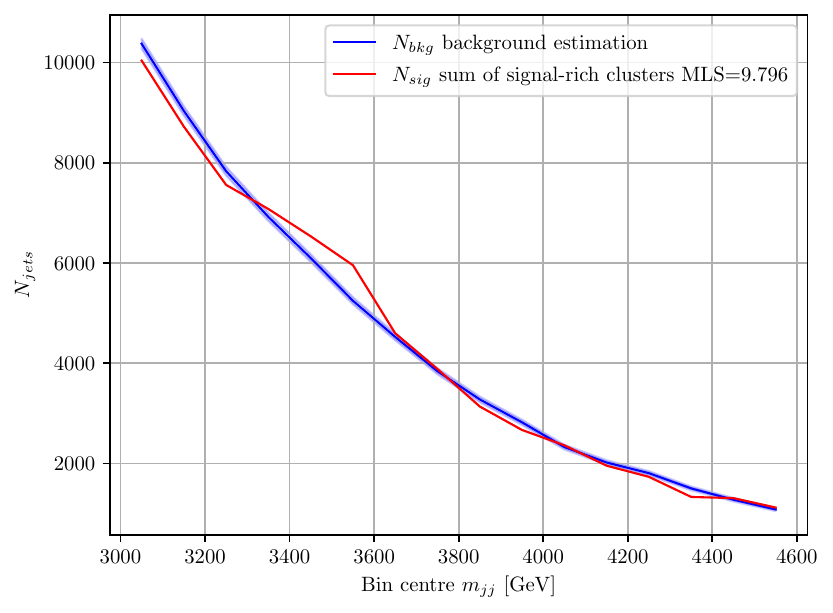}
  \caption{Curves corresponding to the sum of signal-rich and signal-poor spectra in Fig.~\ref{fig:norm_subtr_std}. The blue signal-poor curve is rescaled to have the same total jet number as the signal-rich curve. The coloured region around the blue curve is a $\sigma_{\text{bkg}, b}=\sqrt{N_{\text{bkg},b}}$ Poisson deviation after recaling used to compute MLS.}
  \label{fig:comb}
\end{figure}

\paragraph{Signal-rich and signal-depleted regions:}
After the selection, we combine the non-normalised distributions corresponding to our selected signal-rich clusters. This results in a signal-rich spectrum $N_{\text{sig+bkg},b}$ with an example shown in red in Fig.~\ref{fig:comb}. 

The remaining cluster spectra are combined to form a signal-depleted spectrum $N_{\text{poor},b}$. The estimate of the background is then constructed by normalising it to the same total entries as in signal-rich spectrum, namely $N_{\text{bkg},b}=N_{\text{poor},b}\frac{\sum_b N_{\text{sig+bkg},b}}{\sum_b N_{\text{poor},b}}$. It is shown in blue in Fig.~\ref{fig:comb}.

\paragraph{Test statistic:}
As previously discussed, to test the significance of an observed excess we use the simple maximum local significance, as defined in Equation~\ref{eq:poisson}.
It may occur that no cluster is selected as anomalous. In this case we assign a default value of 0, in order to show good agreement with the null-hypothesis expectation. 
This is similar in motivation to setting the value for an observed deficit in events to zero.
Following the discussion in Subsection~\ref{sub:bump_hunt} for the calibration process, we construct 3,900 pseudo-experiments using bootstrap resampling on 1~million background events. The distribution of the test statistic is discussed in App.~\ref{app:calib_dist}.

\paragraph{Ensembling:}

Different initialisations lead to a broader distribution over the final test statistic obtained with cluster scanning.
In order to obtain a final value for the test statistic, the cluster scanning method is performed 15 times with independent initialisations. The mean of the test statistic from all the runs forms the final ensembled test statistic. The distribution of this statistic is presented in app. \ref{app:calib_dist}.

\subsection{Discussion}
\label{sub:discussion}
As we can see, CS follows the general bump hunt strategy, but introduces novel approaches for the first two steps of this strategy. First of all, CS selects the most anomalous looking clusters to define the signal-enriched region, and constructs a background estimate from the rest of the clusters. Notably though, this selection is completely data-driven and does not target a specific family of signal models. However, CS relies on a set of assumptions that fundamentally differ from those commonly used in other anomaly detection approaches.

\paragraph{Search for overdensity instead of outliers:}
Most anomaly search methods like Autoencoders~\cite{2021AEanomalies} and SVDDs~\cite{aarrestad2022darkMachines} rely on outlier detection, namely, identifying the data instances that lie in a region of very low probability density or outside the support of the ``normal'' distribution. Notably, while all normal events share similar characteristics and exhibit easily recognisable trends, anomalous data, such as defects or fraud, can differ in numerous ways and are thus given a wide prior. Although model-agnostic searches should accommodate a wide range of possible anomaly models, it is usually assumed that a signal is produced by only one or a few unknown BSM process. Thus, all anomalous events have many features in common and exhibit some similarity to SM events, as any new particle must radiate and decay into SM particles to be detectable. 

Therefore we use the localisation of anomalies in both low-level (e.g.\ jet images) and high-level variable (e.g.\ $m_{jj}$) space as the first main assumption of the CS method. Localisation of anomalies in low-level variable space means that only a few out of all clusters contain a fraction of anomalies much higher than the rest of the clusters. This way clustering plays a role of data-driven binning in low-level variable space. Localisation in $m_{jj}$ gives us a possibility to distinguish these anomaly rich clusters from the rest, namely, by searching for an overdensity in $m_{jj}$ in one cluster spectrum compared to all others. Thus, CS is able to select a signal-rich region of events by leveraging the assumption of signal being localised rather than consisting of outliers.

Although semi-supervised methods based on CWoLa (see Refs.~\cite{salad, cathode, curtains, feta, curtainsf4f, buhmann2023phase,Sengupta:2023vtm}) and density estimation methods are also sensitive to overdensities, they usually require construction of a background template, which until recent developments~\cite{buhmann2023phase,Sengupta:2023vtm} was preferably constructed for a smooth distribution of low dimensionality, typically using a few high-level observables. In this publication, we show that CS is able to draw significant improvement from a high-dimensional distribution of low-level jet observables. In this way, it can be considered less signal-model dependent than the methods that rely on hand-crafted high-level observables. 

\paragraph{Assume cluster mass independence instead of smoothness:}

CS proposes a solution to the second step of the analysis, namely, it estimates the form of the background by combining the signal-depleted clusters. In this way, we do not rely on any assumptions on smoothness or on a particular functional form of the background-only spectrum in $m_{jj}$, which are heavily relied upon by most other bump hunt methods, such as the global functional fit mentioned in Subsection~\ref{sub:bump_hunt}, SWIFT~\cite{SWIFT} and even Gaussian processes~\cite{frate2017modeling}. 

Instead, the second main assumption of CS is that in the background-only case, the assigned cluster centroid is approximately independent from $m_{jj}$. Ideally, we would want the distribution of background events over $m_{jj}$ in each cluster to be identical within statistical uncertainty, such that the probability of a jet belonging to a cluster and having a specific mass factorises, $p(i, m_{jj})=p(i)p(m_{jj})$, or at least that the correlation is weak. This would minimise the rate of incorrectly identified signal-enriched clusters. 
In practice, although Fig.~\ref{fig:50_clusters_nrm} shows that the distributions all follow a similar trend, there are still some systematic deviations. These are a result of the finite width in $m_{jj}$ of the training window and slight correlations between the distribution of the jet constituents and $m_{jj}$ arising from the transverse momenta of the two non-resonant jets depending on $m_{jj}$. 
Therefore, for the selection of the clusters, we estimate the uncertainty separately for each experiment and for bin based on the sample of our $k$ cluster spectra values. This uncertainty estimate includes both, statistical Poisson fluctuations and systematical uncertainty from mass dependence (see discussion in App.~\ref{app:gaussian_dist}).

Unlike in methods with sliding window approach~\cite{SWIFT, curtains}, in CS the fit only needs to be performed once \footnote{Training and evaluating CS using a sliding window approach was considered; however, the resulting spectra exhibited abrupt discontinuities due to relatively low statistics in the high $m_{jj}$ bins, making them unsuitable for further analysis.}. Moreover $k$-means clustering is a simple classical algorithm that typically requires less training than deep learning approaches, making CS a relatively fast analysis method.  This is important in the  context of the ensembling and calibration, which both require a large number of analysis iterations, and are thus a notable obstacle to incorporating deep learning in HEP analysis under the constraint in computing resources. Fast analysis is also advantageous for testing its efficiency for simulated BSM events in order to produce the exclusion limits (see the RECAST~\cite{Cranmer_2011} framework). Moreover, CS avoids other disadvantages inherent to sliding window approaches, such as limited search range due to the definition of the sidebands and the need to optimise sideband and signal window widths.

\subsection{Idealised CS}
\label{sub:idealised}

Despite choosing a narrow $m_{jj}$ window to reduce mass dependence systematics, the variables that we use for clustering are in general not independent of $m_{jj}$. 
Thus, we observe the background-only spectra of some clusters do not just statistically fluctuate around the expected shape of the background, but exhibit some degree of smooth mass sculpting. This affects the performance of the method by introducing false positives at the cluster selection stage. 
We expect that this may be partially remedied by a more sophisticated method of selecting anomalous clusters or a better background estimate, both of which would rely on further assumptions. These studies are outside the scope of this publication. However, to give an upper bound on the performance one may achieve with such improvements we propose an idealised version of clusters scanning.

Idealised CS version requires us to modify the distribution of the jets between the clusters. First, we count the numbers of jets that fall into each cluster in the first $m_{jj}$ bin. If no mass dependence were present, the fractions of QCD jets in each cluster $x_{\text{QCD}, i, b} = N_{\text{QCD}, i, b}/\sum_i N_{\text{QCD}, i, b}$ should be independent of bin number $b$ within statistical uncertainties. To simulate this case in all the consecutive bins except the first we distribute the QCD jets in these bins among clusters using a multinomial distribution with weights equal to the fractions obtained in the first bin $x_{\text{QCD}, i, b}=x_{\text{QCD}, i, 1}$, thus generating cluster spectra that follow the original background spectrum with statistical fluctuations, i.e.\ the case with no mass dependence. The signal jets are distributed as before according to which cluster they belong to, such that the fractions of $Z'$ jets may differ between different bins. This is done because we assume that only the background is distributed roughly proportionally between clusters, which is equivalent to assumption 2, but not the signal.

This distribution of jets creates idealised cluster spectra for each clustering, and the rest of the algorithm remains unchanged.

\section{Results}
\label{sec:results}
As a proof of concept we perform an analysis applying CS and global fit based bump-hunting with the above mentioned hyperparameters to the LHCO R\&D dataset with different amounts of signal injection, given in figures either as an absolute number of injected events $\epsilon$ or as a signal to background ratio $S/B$ of events in the considered [3000, 4600]\,Gev $m_{jj}$ region. 
For each pseudo-experiment with signal injection we calculate the significance $Z$ as discussed in Subsection~\ref{sub:significance_ev} using the calibration test statistic distribution. For each signal injection level we run 100 pseudo-experiments with bootstrapped background data and randomly sampled signal events. 
As a reference for the significance and its statistical variation for each contamination level, we report the median significance of these pseudo-experiments and 0.25 and 0.75 quantiles. We define the ratio between the median significance provided by CS to the significance of a baseline method as the significance improvement~(SI).
We also quote the relation between the number of events needed to obtain a $3\sigma$ evidence in each analysis strategy.

\begin{figure}[h]
    \centering
    \begin{subfigure}{0.49\textwidth}
        \includegraphics[width=\textwidth]{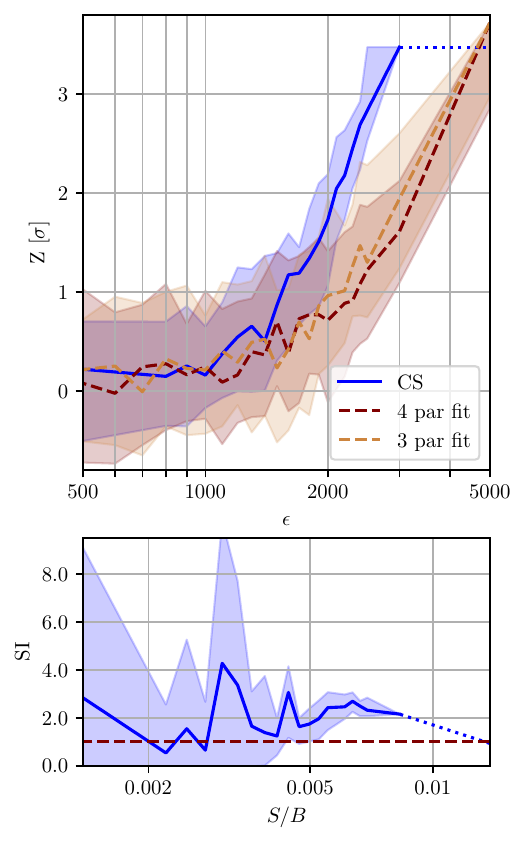}
        \caption{ }
        \label{fig:main}
    \end{subfigure}
    \begin{subfigure}{0.49\textwidth}
        \includegraphics[width=\textwidth]{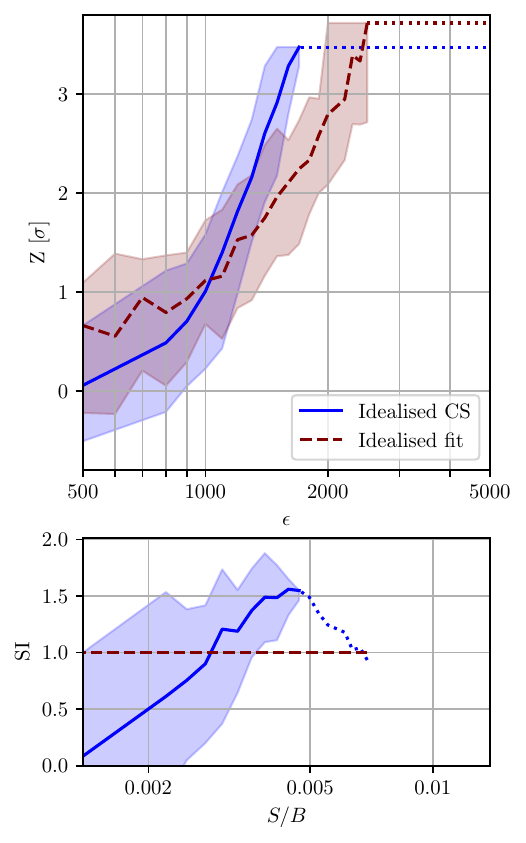}
        \caption{ }
        \label{fig:main_idealised}
    \end{subfigure}
    \caption{Upper figure: Median and the quartile bounds of global significance of the signal contaminated pseudo-experiments as a function of the number of signal events $\epsilon$ injected shown for the (a) realistic and (b) idealised analysis methods. The dotted lines mark lower bounds, as there was not enough statistics to access higher significance levels. Bottom figure: Significance improvement of the CS method compared to the 4-parameter fit  for the (a) and the significance improvement of idealised CS compared to idealised fit  for the (b) as a function of the signal-to-background ratio $S/B$.}
\end{figure}

Figure \ref{fig:main} shows how the global significance of CS and the parametric fit-based methods depends on the signal contamination in the pseudo-experiment. It characterises the performance of these realistic analysis strategies, which do not use any truth information for the evaluation of the test statistic, thus including all the systematical uncertainties coming from partially fulfilled assumptions needed for the respective method. 

We observe that although 3- and 4-parameter fits give approximately the same results, CS outperforms them by a significant margin in the region from 1500 to 4000 signal events. 
Beyond 3000 signal events, the significance yield from CS is limited by the number of bootstrap pseudo-experiments in the calibration set, but the lower bound on its significance still remains substantially higher than the significance of the parametric fits. 
This is the most interesting region as there the transition between non-significant signal (below $1\sigma$) and new physics evidence (above $3\sigma$) takes place.  Looking at the lower subplot in Fig. \ref{fig:main} we see that CS gives us an SI of 2 and higher on the majority of regions of interest. 
We can also see that CS produces a 3$\sigma$ evidence for only 61\% of the events needed to obtain this evidence with the parametric fit. This shows that although both suffer from fit and assumption induced systematic uncertainties, CS has a clear advantage over parametric fitting procedures. 

Above we have also described the idealised version of the cluster scanning method and in Appendix~\ref{app:global_fits} the analysis with an idealised background fit. Both methods rely on event labels to remove systematic uncertainties introduced by the limitations of the assumptions of our methods and to make the background estimate in both cases close to the true background, with only statistical fluctuation taken in consideration. This is done to separate the influence of additional information, namely the low-level observables used in the analysis, from the systematic uncertainties introduced by the fits, and to construct the upper bound on the performance of our methods.

Figure \ref{fig:main_idealised} shows how the global significance of both idealised methods depend on the signal contamination in the pseudo-experiment. It can be seen that one needs substantially less signal for it to be significant in the idealised methods compared to the realistic methods, as it removes false-positives induced by systematic uncertainties in the fits. Still, we see that the idealised CS outperforms the idealised functional fit on the majority of the interval between 1000 and 2200 signal events. Looking at the lower subplot in Fig.~\ref{fig:main_idealised} we see that by using CS in the region of interest we gain a significance improvement factor of up to 1.5. We can also see that CS produces a $3\sigma$ evidence with only 69\% of the events needed to gain this evidence with the idealised fit. This shows that in the case of negligible systematic uncertainties, CS gives an improvement over any smooth fit as, in addition to just using information from $m_{jj}$, it also makes use of the low-level event information. From the difference between idealised and non-idealised CS we can see that there is some room for improvement of CS to reduce the false positive rate, and improve the analysis efficiency.

\section{Conclusions and outlook}
\label{sec:conclusion}
This paper is a first proof of concept for the cluster scanning anomaly search method, which is designed to search for resonant overdensities on the distribution of an observable using clustering techniques in auxiliary observables. 

We found that it outperforms the widely used bump-hunting method, which relies on the functional background fits, in several metrics relevant to an analysis. In the transition region, where the benchmark algorithm achieves 1$\sigma$ to 3$\sigma$ significance, CS improves the result by a factor of 2 or more for the realistic case, or by a factor of 1.5 for the idealised comparison. This reduces the number of signal events required to produce a $3\sigma$ significance by a factor of 0.61 in the realistic case and by a factor of 0.69 in the idealised case. The former factor of improvement should be expected in a real application. We also discuss the comparison of cluster scanning with other anomaly detection algorithms in Subsection~\ref{sub:discussion}, outlining its advantages and limitations. 

The CS method should not be seen as a direct competitor to background fitting methods, but rather as a complementary approach that relies on a different set of assumptions about the nature of the anomaly and the background distributions, which are not well known. 

There remains a large unexplored field of potential extensions and improvements to this method or synergies with other methods. Straightforward follow-up studies can explore the use of clustering methods other than $k$-means. One can look for other ways of selecting the anomalous clusters, alternatives to the one proposed in Subsection~\ref{sub:algo}, that would rely on different assumptions. For example, one can require that all anomalous clusters are neighbors in the space of clustered inputs. One can also unify the assumptions of CS and functional fits to produce separate background estimates for each of the clusters separately, greatly reducing the $m_{jj}$ dependent systematic uncertainties.

CS could benefit from using features developed by other algorithms that have already been optimised for other tasks, such as flavour tagging, or even using unsupervised learning for feature extraction.

Furthermore, since many other ML approaches to improve sensitivity in model-independent searches rely on a bump hunt for the final statistical analysis, CS could also be used to further enhance sensitivity. This could be of particular interest when the background distribution is no longer well described by simple, smoothly decreasing functional forms.

\acknowledgments

The authors would like to acknowledge funding through the SNSF Sinergia grant CRSII5\_193716 ``Robust Deep Density Models for High-Energy Particle Physics and Solar Flare Analysis (RODEM)''
and the SNSF project grant 200020\_212127 ``At the two upgrade frontiers: machine learning and the ITk Pixel detector''. The research of MK is supported by the DFG under grant 396021762 - TRR 257: Particle Physics Phenomenology after the Higgs Discovery.

\section*{Code availability}

The code used to produce all results presented
in this paper is available at \url{https://github.com/IvanOleksiyuk/jet_cluster_scanning} .

\section*{Appendix}
\appendix
\section{Idealised fit and n-parameter fit pseudo-analysis}
\label{app:global_fits}
In general, if a distribution $\mathcal{H}_b$ of a background events is perfectly known a-priory, given a binning for this distribution, one can calculate the expected number of events in each bin. Being provided directly from the true underlying $\mathcal{H}_b$, this background estimate will on average provide the most efficient tests to discriminate samples drawn from $\mathcal{H}_b$ from samples drawn from an alternative hypothesis $\mathcal{H}_{b+s}$ with signal, compared to any other estimate of the expected background for these samples. Hence we call the expectation from $\mathcal{H}_b$ an "idealised fit".

As discussed extensively in section \ref{sub:bump_hunt}, estimation of the expected background is only one step of the analysis. To create a benchmark, we do pseudo-analysis on LHCO R\&D dataset represented by spectrum $N_{\text{bkg}, \text{orig}, b}$ using the other choices defined in section \ref{sub:bump_hunt}. Namely, we generate pseudo-experiments by bootstrap resampling the events from $N_{\text{bkg}, \text{orig}, b}$ and add a number of signal events if needed. The "idealised" background estimation for every pseudo-experiment is equal to $N_{\text{bkg}, \text{orig}, b}$ itself (as the samples were generated with these expected values). Following the discussion in subsection \ref{sub:test_stat} we use MLS test statistic between this estimate and the generated pseudo-experiments, to generate null-hypothesis test statistic distribution and its value for signal contaminations and thereafter estimate the significance. Depending on the number of doped signal events, the median and quartile region significance given by this test is provided in the main text in Fig.~\ref{fig:main_idealised}.

Unfortunately the background model is usually unknown, so for each experimental sample the background should be estimated in some less precise way relying on weaker assumptions. 

As a realistic benchmark to our method we explore how sensitive the analysis is using global n-parameter functional to the kind of signal presented in LHCO R\&D dataset. We use the binning with 16 bins defined in subsection \ref{subsec:bkg} and count the number of background events in each bin to get an original background spectrum $N_{\text{bkg}, \text{orig}, b}$. 

For all the fits in this studies, we use Trust Region Reflective nonlinear least squares fitting algorithm implementation from \textsc{Scipy} python package \cite{2020SciPy-NMeth}.  The chosen bins generally contain more than 5000 counts, so the Poisson distributions of these counts can be well approximated by a Gaussian distributions with the variances equal to the bin counts. Using variances to scale the summands in the least squares objective we make it equivalent to the maximum likelihood objective for this setup.

First, we fit our 3- and 4-parameter functions to the spectrum to see if the fit is valid. 
Resulting fits with 13 and 12 degrees of freedom score $\frac{\chi^2_{3-par}}{n_{dof}}\approx1.201$ and $\frac{\chi^2_{4-par}}{n_{dof}}\approx1.338$ that correspond to p-values of $0.275$ and $0.182$ which signify validity of these fits. 

Unlike the CS method that doesn't generally rely on the smoothness of the background, global n-parameter takes it as the main assumption, so as $N_{\text{bkg}, \text{orig}, b}$ already has some statistical fluctuations a distribution resampled from it will have even larger statistical fluctuations than the ones expected for Poisson distribution. To simulate the proper scale Poisson fluctuations in the chosen region for our pseudo-experiments we resample events not from $N_{\text{bkg}, \text{orig}, b}$ but from the best possible fit. This also negates the systematic error from null-hypothesis not corresponding to the empirical functional form, so these experiments can be viewed as semi-idealised. In a more realistic cases, the space of functions given by all possible parameter values, does not contain the true form of null-hypothesis distribution and can only yield an approximation of it with limited precision. 
It is usual for fit functions with a small number of parameters, but with increasing number of parameters the function fit problem becomes over-defined and the function can fit the signal bump as well. Experimentally we have observed only insignificant increase in performance when comparing sampling from $N_{\text{bkg}, \text{orig}, b}$ or from the best fit distributions. On top of the resampled background events we add a number of signal events from signal's original distribution when needed.

The initial parameters of the fit in each experiment are chosen to be equal to the optimal parameters of the initial fit discussed above, so that one gets an "idealised" background fit if no optimisation is done. However, because of the statistical fluctuations and/or added signal contamination, the maximisation of likelihood results in a different set of parameters for this functional form. This error of background mismodeling under its statistical fluctuations and addition of the signal is exactly the type of error we want to demonstrate with this pseudo-analysis. 

The results of such analysis for different signal contamination is given in Fig.~\ref{fig:main}. We can see that the 3-parameter fit provides a slightly better result than 4-parameter fit as the latter has more flexibility to overfit the signal and the statistical fluctuations. This is so because the samples are drawn from 3- and 4-parameter functions with fixed parameters themselves. 
If we were to sample from other distribution the error coming from mismatch in true end expected functional forms may switch this ordering but it will reduce both performances. 
Therefore, the curves shown in Fig.~\ref{fig:main} are upper limits of these realistic n-parameter fit analyses achievable only when the true distribution is described by one of the functions in the chosen parameterised space.

\section{Sparsity of the jet images}
\label{app:sparse}
\begin{figure}[H]
    \centering
    \includegraphics[width=0.8\linewidth]{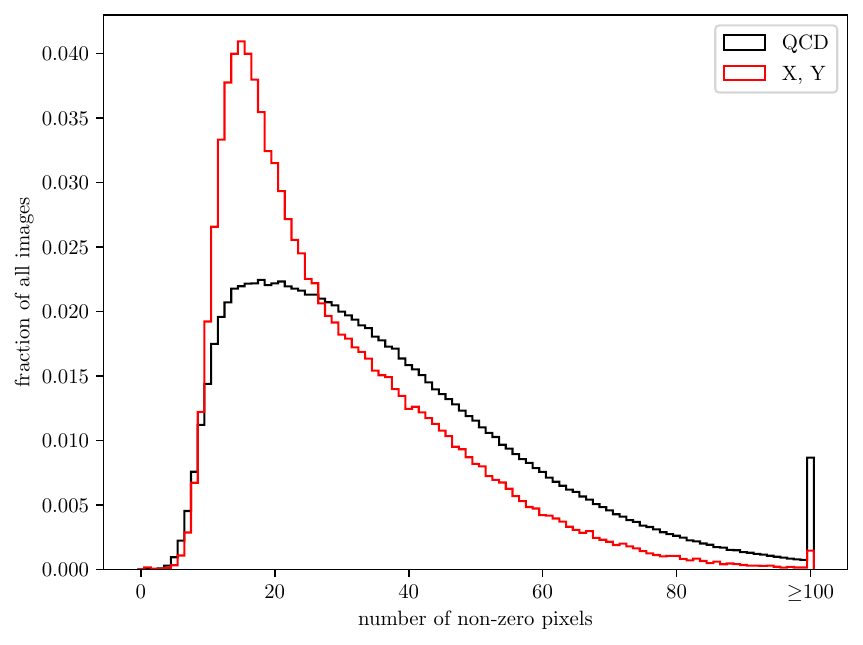}
    \caption{Distribution of images in QCD and top datasets from top-tagging task vs the number of non-0 pixels in them.}
    \label{fig:population}
\end{figure}

Fig.~\ref{fig:population} show that the jet images are very sparsely populated ususally having less tha 100 non-zero pixels per 1600 pixels total.

\label{app:sparce}

\section{Hyperparameter selection and motivation}
\label{app:hyper}

In this appendix we give motivation for every not yet discussed choice of hyperparameter in our pseudo-analysis. All the hyperparameter suggestions are done in an unsupervised way coming from general assumptions about signal and background and are not optimised using the truth information from LHCO R\&D data. As such the levels of significance improvements may be further increased by performing a dedicated parameter scan for a specific application, however, we recommend to follow the same reasoning when applying CS in other analyses.

\paragraph{Training region:}
Training on the full spectrum would likely result in each cluster corresponding to a specific mass region, thus the background spectrum for each cluster would not be close to the original mass spectrum. Therefore we perform clustering in a narrow mass window [3000, 3100]\,GeV. We choose this window as it lies in the studied region defined in subsection \ref{sub:sig_en} and has the largest statistic of all other 100\,GeV windows. 

\begin{figure}[h]
    \centering
    \includegraphics[width=0.8\linewidth]{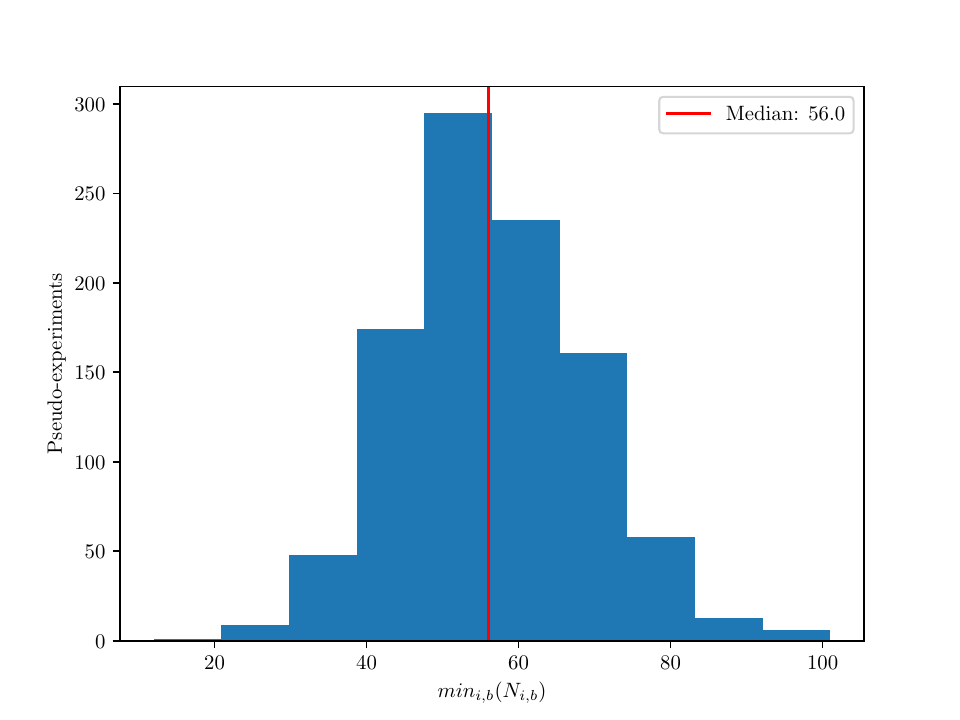}
    \caption{Distribution of the number of jet image counts in the least populated bin of the least populated cluster in each of the 1000 random background only runs of the CS algorithm on backround only data.}
    \label{fig:min_count}
\end{figure}

\paragraph{Number of clusters:}
The most important parameter we had to choose is the number of clusters $k$. Two factors play the key role in this choice. On one hand, the number of clusters has to be as large as possible to better narrow down the anomaly-rich region. On the other hand, for a given number of events in the evaluation region and the binning of this region one has to take the number of clusters sufficiently small so that the least populated clusters in the smallest $m_{jj}$ bin $\underset{i, m_{jj}}{min}(N_{i}(m_{jj}))$ still has enough statistics  for a meaningful statistical analysis. We assume that $\underset{k, m_{jj}}{min}(N_{i}(m_{jj}))=O(50)$ should be sufficient. Using a coarse search, we determine, that for our choice of binning and overall statistic at hand choosing $k=50$ gives a good trade-off as it has a median of 55 events in smallest cluster-bin and it goes below 20 only 1 time in 1000 pseudo-experiment runs, as can be seen in Fig.~\ref{fig:min_count}.

\paragraph{Batch size:} Scikit-learn \cite{scikit-learn} documentation states that the parallelisation is performed on all available $N_{cores}$ computing cores if the batch size is $N_{cores}\cdot256$ or larger. We performed all computations with 8 core parallelisation, thus the natural choice of a batch size was 2048. It is also important to maintain the batch size much larger than the number of clusters to ensure faster convergence. 

\paragraph{Outlier fraction:} 
To quantify the performance of our method we introduce the signal fraction improvement score (SFI) that characterises a subset $\mathcal{S}$ of the events in evaluation set $\mathcal{E}$ by the relative increase in the signal to background ratio
\begin{equation}
    SFI(\mathcal{S}) = \frac{N_{\text{sig}}(\mathcal{S})}{N_{\text{QCD}}(\mathcal{S})} \frac{N_{\text{QCD}}(\mathcal{E})}{N_{\text{sig}}(\mathcal{E})}.
\end{equation}
Our main assumption is that the signal is distributed in clusters unevenly and there only several clusters have a significantly large SFI. To put it in numbers, we assume that not more than 20\% of clusters have SFI of 2 or more. Following this assumption we choose the outlier fraction of 0.2 for outlier robust estimators. 
This is an ad.\ hoc prior assumption about the data at hand, and it has to be made prior to analysis and has no way to be validated without knowing the truth lables.
Still we can show that this assumption is satisfied in our case with a margin for the pseudo-experiment shown on all the figures of  section \ref{sec:method}. 
5000 signal events were giving an overdensity on the original spectrum that was not identifiable as a deviation from smooth background by human eye (without knowing the background truth), but in Fig.~\ref{fig:50_clusters_nrm} one can easily notice two spectra with a significant bump around 3.5\,TeV that stand out of the crowd of other spectra. Unsurprisingly these two spectra have SFIs of 9.1 and 8.9. 
Three more clusters also have a visible overdensity at this position possessing SFIs of 6.3, 5.6, 4.4. 
In total, exactly 8 clusters have $SFI>2$. Still as we will see later only 3 of these clusters have a signal significant enough to be selected as anomalous, showing that our assumption is quite conservative in its limit and either the threshold $SFI$ can be increased or the percentage of clusters to path the threshold reduced for it to still remain a valid assumption. 
Runs of the analysis on other (pseudo-)experiments behave in the similar manner.

\paragraph{Cluster selection threshold $\theta$:}

First of all, we use the threshold only for positive deviations as we only search for excesses of events. Apart from the signal-rich outlier clusters the threshold can be passed by signal poor clusters, but only with an expected false positive rate of $1-(1-\text{p-value}_{\mathcal{N}(0, 1)}(\theta))^{n_{bins}}$. Then for large enough thresholds the average number of false positives can be estimated as $k\cdot n_{bins}\cdot \text{p-value}_{\mathcal{N}(0, 1)}$. Higher thresholds result in lower false positive and lower true positive rates. To retain the sensitivity for statistically small signal we choose to use $\theta=3$ that will result in approximately $50\cdot16\cdot0.00135=1.08$ signal poor cluster being assigned a false positive label on average. Fig.~\ref{fig:norm_subtr_std} shows 4 clusters being chosen using this threshold. Three of them have an overdensity at 3.5\,TeV and one does not, implying that it is a likely false posive. 

\paragraph{Ensemble size:} We recommend to take the ensemble size as high as possible, for given computation resource constrains to reduce the width of the test statistic distribution (see appendix \ref{app:calib_dist}).

\begin{table}[H]
    \centering
    \begin{tabular}{|c|c|c|}
        
        \hline
         Parameter & value & motivation \\
         \hline
         \hline
         k & 50 & $\underset{i, b}{min}(N_{i,b})=O(50)$ with binning below \\
         \hline
         mini-batch & 2048 & $N_{cores}\cdot256$, must be $\gg k$ \\
         \hline
         Training region & [3, 3.1]\,TeV & narrow mass window with high statistic \\
         \hline
         Evaluation region & [3, 4.6]\,TeV & the n-parameter fit is applicable  \\
         & & excluding low statistic regions \\
         \hline
         Bin width & 100\,Gev & broad enough to have sufficient statistics in each bin \\
         \hline
         outlier & 0.2 & consistent with assumption \\
         fraction $f$ & & on the maximum number of signal clusters\\
         \hline
         Cluster selection & 3 $\sigma$ & low enough to let trough many true positives  \\
         threshold $\theta$ & & but high enough to filter most false positives \\
         \hline
         Test statistic & MLS & simple and specialised for local excesses \\
         \hline
         Default TS & 0 & minimal test statistic possible \\
         \hline
         Ensemble size & 15 & As large as possible realistic compute limitations \\
         \hline
    \end{tabular}
    \caption{Summary of the hyperparameters used in cluster scanning.}
    \label{tab:hyp}
\end{table}

\section{Distribution of cluster scanning bin entries}
\label{app:gaussian_dist}
\begin{figure}[h]
  \centering
  \includegraphics[width=1\textwidth]
  {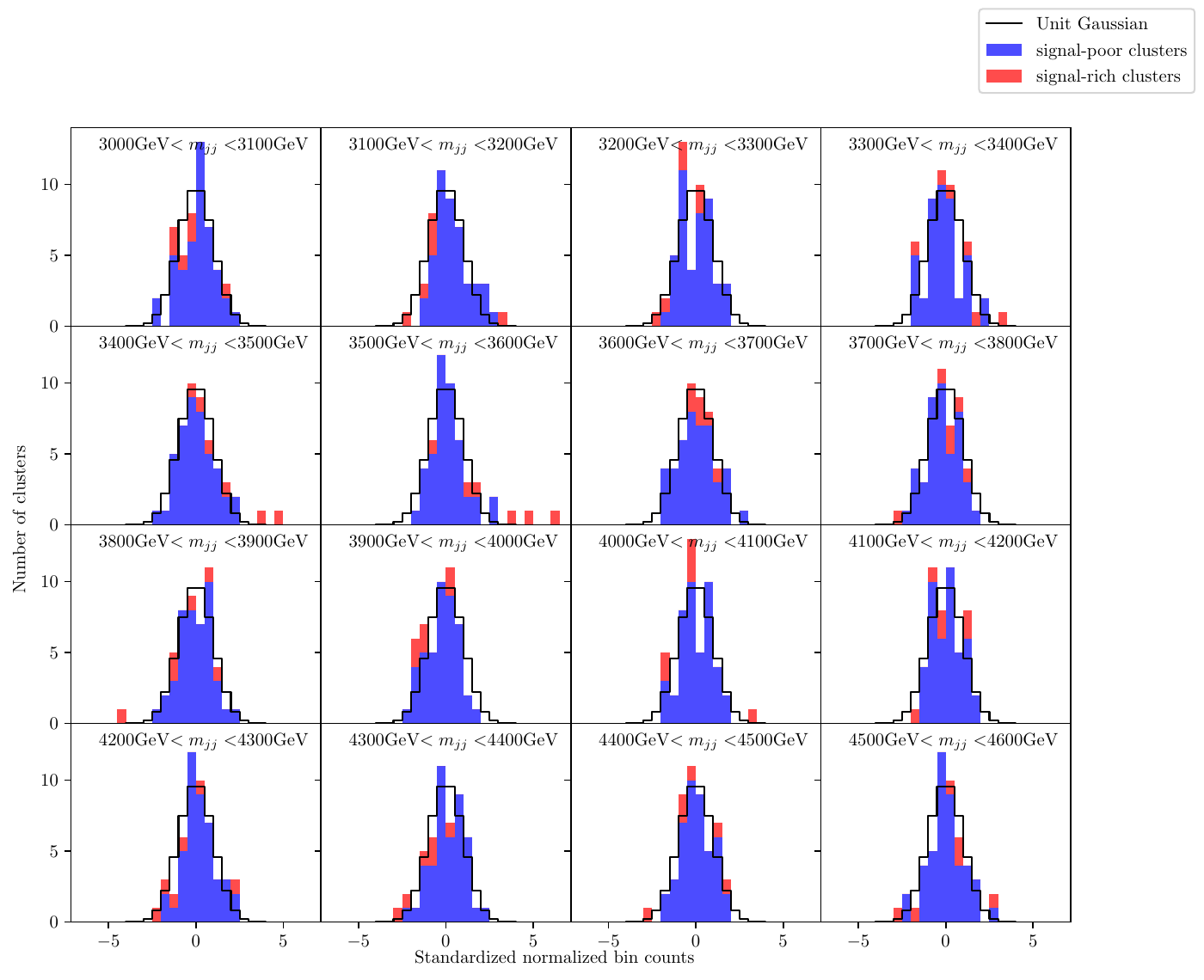}
  \includegraphics[width=0.75\textwidth]
  {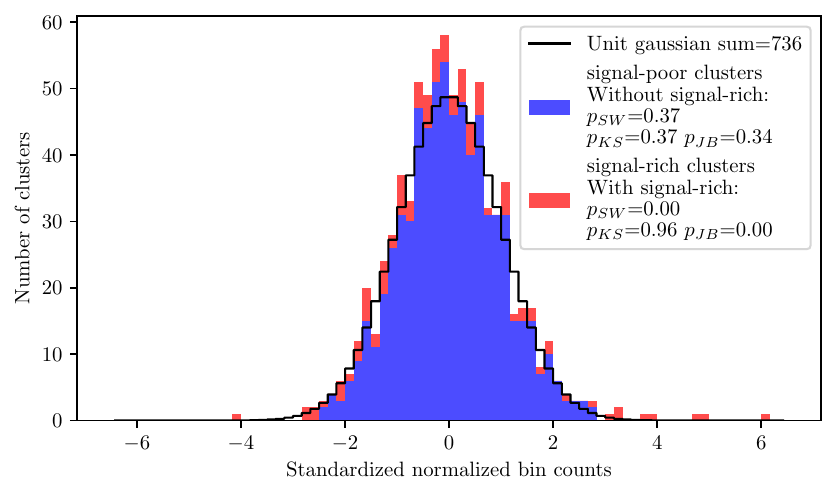}
  \caption{Top: a histograms of cluster spectra for each bin in Fig.~\ref{fig:norm_subtr_std}, depending on the deviation from the robust mean measured in robust standard deviations. Bottom: sum (marginal) of all the histograms at the top compared to expected bin counts of a Gaussian with a sum of 736 count (all the signal poor counts).}
  \label{fig:gaussianity_check_all_bins}
\end{figure}

The assumption on the Gaussianity of cluster spectra in each bin can be shown to be valid by robustly standardizing the cluster counts in each bin (see App.~\ref{app:robust_estimators}) and checking if these distributions match $\mathcal{N}(0, 1)$. The upper part of Fig.~\ref{fig:gaussianity_check_all_bins} visually demonstrates that the distribution of cluster spectra in each bin in Fig.~\ref{fig:norm_subtr_std} matches the Gaussian model, except in bins $3.4 \text{\,TeV}<m_{jj}<3.5 \text{\,TeV}$ and $3.5 \text{\,TeV}<m_{jj}<3.6 \text{\,TeV}$, where many outlier clusters are found due to the presence of signal. Fifty samples are usually not sufficient to determine whether the distribution is Gaussian or not, but by marginalizing (summing up) over 16 bins, we obtain 800 samples in total. The lower plot in Fig.~\ref{fig:gaussianity_check_all_bins} shows the said distribution. If we consider only signal-poor clusters, the distribution fits the $\mathcal{N}(0, 1)$ distribution well visually and according to a consensus of three normality tests: Shapiro-Wilk, Kolmogorov-Smirnov, and Jarque-Bera (note that the p-values for all the tests are high). Including signal-rich clusters adds outliers, which is reflected in lower p-values for the normality tests; however, apart from these, it still can be well approximated by a unit Gaussian.

The variance among clusters in each bin depends on both statistical fluctuations and mass-dependent systematic effects. In the case of infinite statistics, all cluster spectra would appear smooth but would vary from one another due to differences in mass sculpting. In the absence of mass dependence, the variation in each bin would be caused solely by Poisson fluctuations (as mentioned in subsection \ref{sub:idealised}, dedicated to idealised CS). In a realistic scenario, these two factors cannot be separated but can be jointly estimated using a robust standard deviation for each bin (see App.~\ref{app:robust_estimators}).

\section{Outlier robust estimators}
\label{app:robust_estimators}
While searching for outliers, it is preferred to use outlier robust estimators for standard deviation (SD) and mean. We define them as follows: given a sample of observations $S=\{x_1, x_2, ...\, x_n\}$ we find a median $med(S)$ (which is itself an outlier robust estimator) of this sample and take a subsample $\Tilde{S_f}$ that is constructed from $S$ by discarding a fraction $0<f<1$ of all samples that have largest absolute distance to this median. 
In this way we have discarded the outliers. After that we construct estimators $\Tilde{\mu}_f=mean(\Tilde{S}_f)$ and $\Tilde{\sigma}_f=SD(\Tilde{S}_f)\cdot g(f)$. If $S$ is a sample from $\mathcal{N}(\mu, \sigma)$ it is obvious that with $\underset{n \rightarrow \infty}{lim}\Tilde{\mu}_f=\underset{n \rightarrow \infty}{lim}mean(S)=\mu$. 
If one takes $S$ from $\mathcal{N}(0, 1)$ and rescales $x_i\rightarrow\sigma x_i$, then both estimators transform as $\Tilde{\sigma}_f\rightarrow\sigma\Tilde{\sigma}_f$ and $SD(S)\rightarrow \sigma SD(S)$ by definition, so both estimators $\Tilde{\sigma}_f$ and $SD(S)$ are proportional to a true $\sigma$ of the Gaussian distribution. 

Both $\Tilde{\sigma}_f$ and $SD(S)$ are independent of $\mu$ and there are no other parameters of the normal distribution for estimators to depend on, therefore for a family of Gaussian distribution estimators $\Tilde{\sigma}_f$ and $SD(S)$ are proportional to each other by some constant factor $g(f)$ in the limit of infinite sample. In other words, adjusting numerically $g(f)=\frac{SD(\mathcal{N}(0, 1))}{\Tilde{\sigma}_f(\mathcal{N}(0, 1))}=\frac{1}{\Tilde{\sigma}_f(\mathcal{N}(0, 1))}$ is sufficient to sattisfy $\underset{n \rightarrow \infty}{lim}\Tilde{\sigma}_f=\underset{n \rightarrow \infty}{lim}SD(S)=\sigma$. So $\Tilde{\mu}_f$ and $\Tilde{\sigma}_f$ are unbiased estimators of $\mu$ and $\sigma$ of a normal distribution, although depending on $f$ they are less efficient than usual non-robust mean and SD.

\begin{figure}[h]
  \centering
  \includegraphics[width=0.8\textwidth]{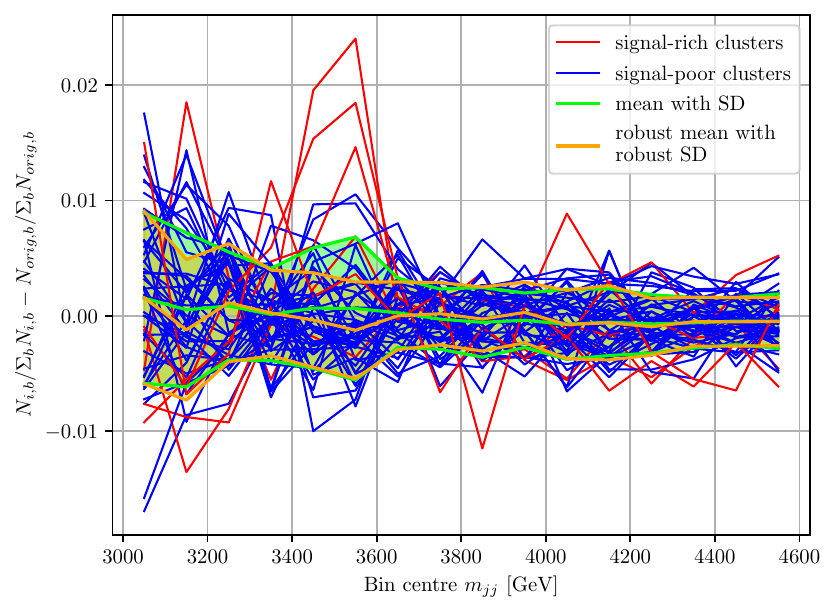}
  \caption{ Normalised spectra with subtracted normalised original $m_{jj}$ spectrum. Amount of signal is 5000. The selection of the anomalous clusters is taken from Fig.~\ref{fig:comb}.}
  \label{fig:norm_subtr}
\end{figure}

Fig.~\ref{fig:norm_subtr} shows us the cluster spectra from Fig.~\ref{fig:50_clusters_nrm} with subtracted normalised original spectrum (which is only needed for better visualisation as this step has no effect on the standardisation).

Fig.~\ref{fig:norm_subtr} also shows the conventional and the outlier robust estimations of mean and SD of the cluster spectra values in each bin. As expected for lower $m_{jj}$ the SD is higher as these deviations is partially caused by the Poisson fluctuations which are proportional to $\sqrt{N_{i,b}}$. We can also see the conventional estimators have a bump around 3.5\,TeV that is induced by our outlier signal-rich clusters, while the robust estimators are unaffected by the outliers.

\section{Calibration distributions}
\label{app:calib_dist}
\begin{figure}[h]
    \centering
    \begin{subfigure}{0.99\textwidth}
        \includegraphics[width=\textwidth]{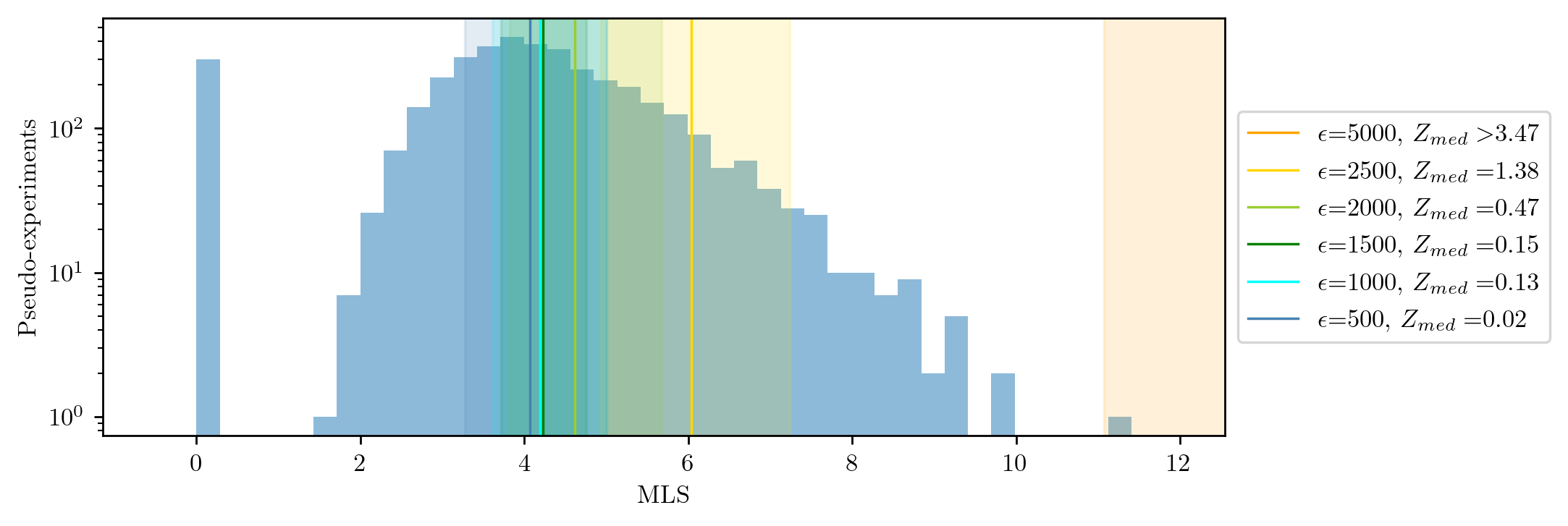}
        \caption{ }
        \label{fig:dist_1m}
    \end{subfigure}
    \begin{subfigure}{0.99\textwidth}
        \includegraphics[width=\textwidth]{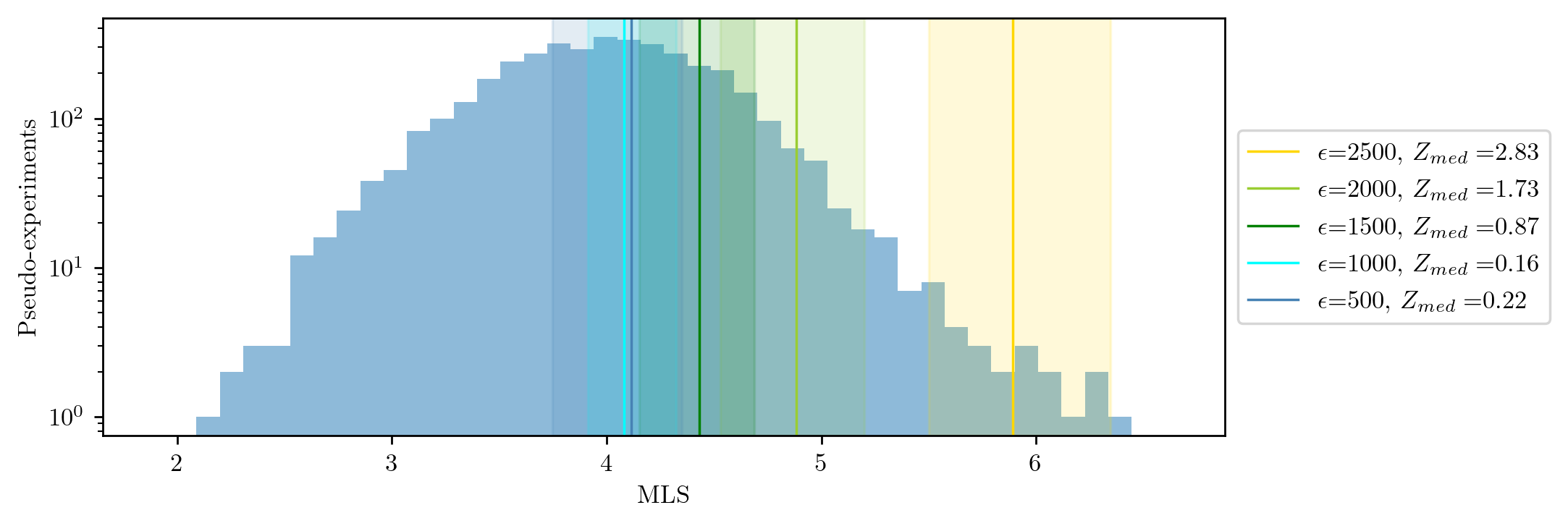}
        \caption{}
        \label{fig:dist_15m}
    \end{subfigure}
    \begin{subfigure}{0.99\textwidth}
        \includegraphics[width=\textwidth]{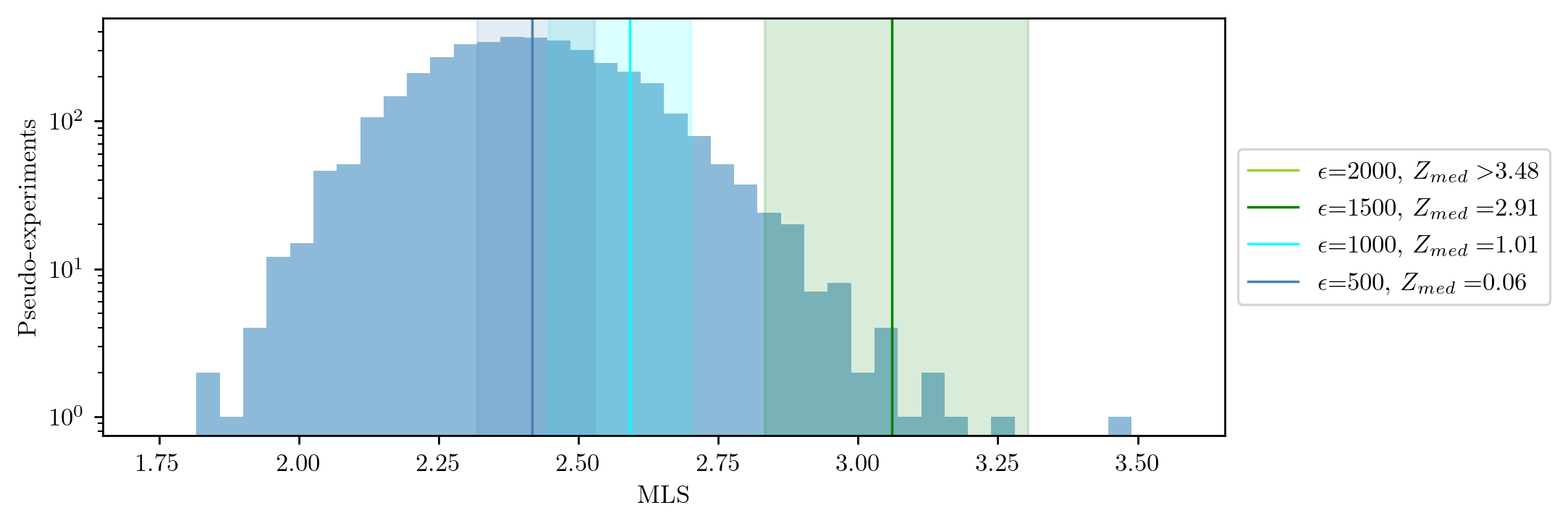}
        \caption{}
        \label{fig:dist_ideal}
    \end{subfigure}
    \caption{Histogram of the CS test statistic for pseudo-experiments with bootstrapped background only samples. Vertical lines and vertical bands show median and region between lower and higher quartiles of test statistics for pseudo-experiments with signal injection. Several signal injection levels are represented by different colours. Panel (a) shows a case with only 1 initialisation of clusters in CS per pseudo-experiment, panel (b) shows a case for ensembling 15 runs of CS with different intialisations per pseudo-experiment and panel (c) shows a case for ensembling 15 runs of idealised CS with different initialisations per pseudo-experiment.}
\end{figure}

The distribution of the test statistics given by CS without ensembling for all background only pseudo-experiments is shown in Fig.~\ref{fig:dist_1m} as a histogram. We see that around 300 of those were assigned test statistic of 0 as they had no clusters selected as anomalous. Other cases where one or more anomalous clusters were selected form a smooth continuous distribution.

The median CS test statistic for 100 signal contaminated pseudo-experiments is represented in Fig.~\ref{fig:dist_1m} by a vertical line, and the vertical band represent the region between the quartiles of such a test statistic sample. For each signal-doped pseudo-experiment we calculate significance as it is described in subsection \ref{sub:significance_ev}. The median significance is quoted in the legend of the figure.

Fig.~\ref{fig:dist_15m} shows the distribution that is analogous to the one in Fig.~\ref{fig:dist_1m}, but with an ensemble of 15 runs of CS algorithm for each pseudo-experiment. 
We notice that the distribution in Fig.~\ref{fig:dist_15m} is significantly narrower than in \ref{fig:dist_1m} which reduces the frequency of background only experiment having large test statistic, thus increasing the sensitivity to signal injection. 
An additional benefit is that the uncertainty region (between two quarterlies) for each signal doping have significantly decreased which is important for lower uncertainty in the analysis on experimental data on the excess significance or on the exclusion limits.

This motivates, that in general the ensemble size should be taken as large as reasonably possible. Our choice of ensemble size 15 together with the number of pseudo-experiments 3900 were dictated by the computing time and storage memory limits as the amount of full CS algorithm iterations is the product of those numbers (excluding the pseudo-experiments with signal injection)

Finally, Fig.~\ref{fig:dist_ideal} shows that for idealised CS without systematics introduced by mass correlations the MLS between our signal spectrum and background estimate is lower. 
Moreover, as expected, it improved the sensitivity of the method to the signal. Obviously this technique cannot be utilised in an actual analysis as jet labels are needed to distribute signal and background jets in a different manner.

\section{Impact of signal in the training region}
\label{app:signal_in_training}

\begin{figure}[h]
  \centering
  \includegraphics[width=0.83\textwidth]{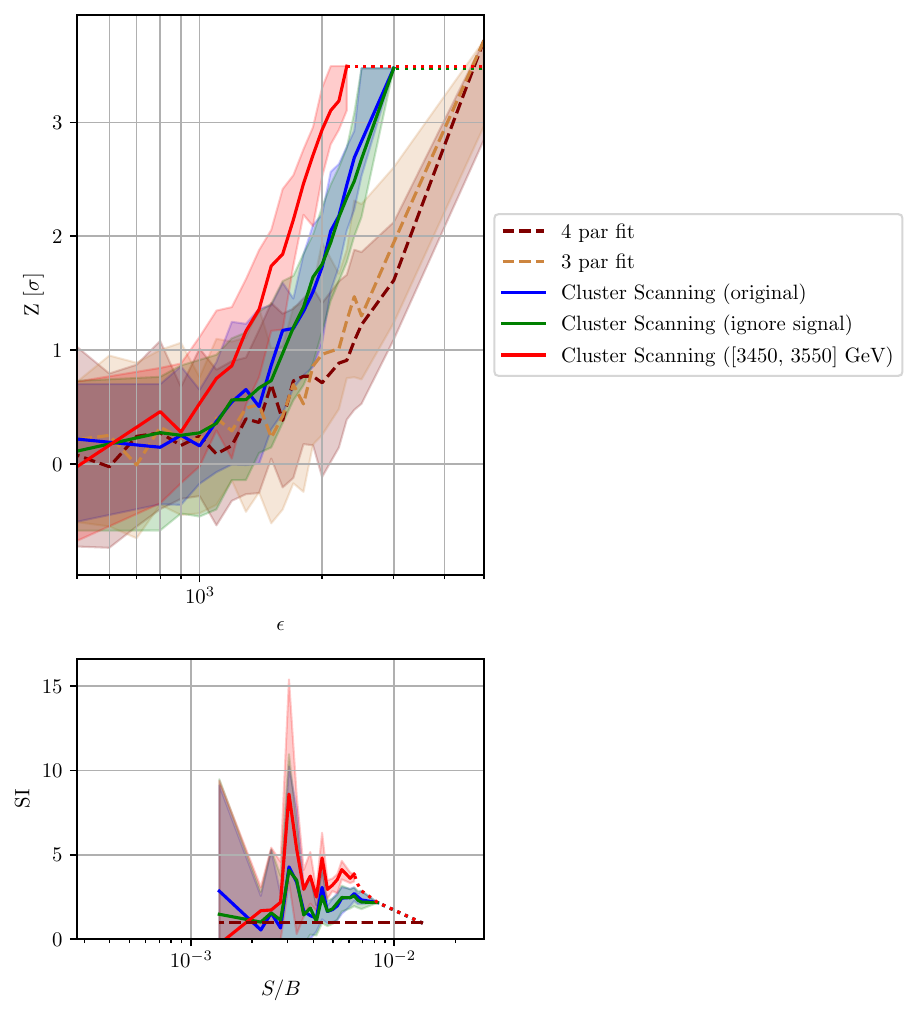}
  \caption{Analogous to Fig.~\ref{fig:main}, with two curves added for the experiments of training without signal and for training in the most signal-rich region of [3450, 3550]\,GeV.}
  \label{app:main_signal_ablation}
\end{figure}

First of all, we retrain all the signal-contaminated experiments such that we do not include the O(100) or fewer signal events while performing k-means fit. We still include the correct number of signal events when evaluating. The performance of this version of CS is demonstrated with the curve labeled as "ignore signal" in Fig.~\ref{app:main_signal_ablation}. It is evident that the two versions have a statistically insignificant difference. We conclude that indeed the original CS version well describes a general and realistic case of having negligible to no signal contamination in the training region.

For the next experiment, we train clustering in the region with the highest signal event fraction, namely [3450, 3550]\,GeV. We run background-only and signal-contaminated pseudo-experiments in this region with all other hyperparameters equal to the values used in main studies. In the case of a non-negligible signal fraction in the training region, the cluster centroids will be attracted to the regions of signal event concentration. We observe this effect, as the resulting signal clusters have a much higher signal fraction as the signal events "pull" the corresponding cluster centers closer, leading to a large increase in the performance of the CS method that is visible in Fig.~\ref{app:main_signal_ablation}. Although in the general case the position of the signal-rich region is unknown, these studies prove that Fig.~\ref{fig:main} and Fig.~\ref{fig:main_idealised} only show the lower bound for the discovery potential of CS.

\bibliography{bibliography}
\bibliographystyle{JHEP}

\end{document}